\documentclass[useAMS,usenatbib]{mn2e}
\usepackage{graphicx}
\usepackage{ifpdf}

\newcommand{\nablab}{\mathbf{\nabla}}

\usepackage{times}

	\usepackage{hyperref}
	\hypersetup{colorlinks=true,citecolor=blue,linkcolor=black,filecolor=black,runcolor=black}

\usepackage{booktabs}

\usepackage{aas_macros}

\ifpdf
	\usepackage{epstopdf}
\fi

\title[Photoevaporation and planet formation]{The interplay between X-ray photoevaporation and planet formation}
\author[G. P. Rosotti et al.]{Giovanni P. Rosotti\thanks{E-mail:
rosotti@usm.lmu.de}$^{1,2,3}$, Barbara Ercolano$^{2,3}$, James E. Owen$^{4}$, Philip J. Armitage$^{5,6}$\\
$^{1}$Max-Planck-Institut f\"ur extraterrestrische Physik, Giessenbachstra\ss{}e, D-85748 Garching, Germany\\
$^{2}$ Excellence Cluster Universe, Boltzmannstr. 2, D-85748 Garching, Germany\\
$^{3}$ Universitats-Sternwarte M\"unchen, Scheinerstra\ss{}e 1, D-81679
M\"unchen, Germany\\
$^{4}$ Canadian Institute for Theoretical Astrophysics, 60 St. George Street, Toronto M5S 3H8, Canada\\
$^{5}$ JILA, University of Colorado and NIST, 440 UCB, Boulder, CO 80309-0440, USA\\
$^{6}$ Department of Astrophysical and Planetary Sciences, University of Colorado, Boulder, USA}

\begin{document}

\date{Accepted 2012 December 28.  Received 2012 December 21; in original form 2012 November 8}

\pagerange{\pageref{firstpage}--\pageref{lastpage}} \pubyear{2012}

\maketitle

\label{firstpage}

\begin{abstract}

We assess the potential of planet formation instigating the early formation of a photoevaporation driven gap, up to radii larger than typical for photoevaporation alone. For our investigation we make use of hydrodynamics models of photoevaporating discs with a giant planet embedded. We find that, by reducing the mass accretion flow onto the star, discs that form giant planets will be dispersed at earlier times than discs without planets by X-ray photoevaporation. By clearing the portion of the disc inner of the planet orbital radius, planet formation induced photoevaporation (PIPE) is able to produce transition disc that for a given mass accretion rate have larger holes when compared to standard X-ray photoevaporation. This constitutes a possible route for the formation of the observed class of accreting transition discs with large holes, which are otherwise difficult to explain by planet formation or photoevaporation alone. Moreover, assuming that a planet is able to filter dust completely, PIPE produces a transition disc with a large hole and may provide a mechanism to quickly shut down accretion. This process appears to be too slow however to explain the observed desert in the population of transition disc with large holes and low mass accretion rates.

\end{abstract}


\section{Introduction}
The evolution and final dispersal of the gas and dust contained in protoplanetary discs surrounding young low-mass stars plays an important role in shaping the properties of potential planets in the system. In particular the disc dispersal time-scale sets an upper limit for the formation of gas giants. It is therefore crucial to formulate a theory capable of describing the disc evolution and dispersal in detail, which is an important input for the planet formation theory.

For most of their lifetime, the evolution of discs is driven by viscosity. Currently, the most favoured mechanism for its origin is the magnetorotational instability \citep{BalbusHawley}, although other mechanisms have been proposed, such as the transport of angular momentum by spiral waves in self-gravitating discs \citep{LodatoGravInstability}. Under simplifying assumptions, the evolution in time of a disc can be described by an analytical solution \citep{lyndenbellpringle}, which was showed to be in rough agreement with the observations \citep{Hartmann98} of the mass accretion rates in ``classical'' T-Tauri discs.

However, the final evolution of protoplanetary discs appears not to be compatible with this picture. Discs have a characteristic lifetime of ~3 Myr \citep{Haisch2001,Mamajek2009,Fedele2010}, as can be shown looking at the fraction of disc bearing young stellar objects (YSO) in clusters of different ages, but, rather than from a homogeneous draining as predicted by pure viscosity evolution, discs seem to have a fast, final stage of clearing from the inside out \citep{2010ApJS..186..111L,2011MNRAS.410..671E,2012arXiv1210.6268K} with a typical time-scale of $10^5$ years. This behaviour has been called ``two-timescale'' in the literature.

``Transitional discs'' are objects believed to have been caught in the act of disc-dispersal and hence to be useful for shedding light on the mechanism responsible for disc clearing. First spotted by spectral energy distribution (SED) observations of YSOs more than two decades ago \citep{Strom89,Skrutskie90}, although only more recently the Spitzer space telescope gave the possibility to study them in detail \citep[e.g. ][]{Calvet05,Espaillat10}, they lack emission at the mid IR wavelengths when compared to ``standard'' disc. This deficit of opacity in the warm dust has been interpreted as the signature of an inner hole. SED modelling is however a rather difficult task, depending on many model parameters that are often degenerate. More recently, the advent of good quality data from sub-mm interferometers has permitted to obtain spatially resolved images of some of these objects, confirming indeed the presence of large cavities \citep{Andrews2011}, sometimes of order of tens of AU. The frequency of transitional disc ($\sim$ 10 per cent of total number of protoplanetary discs) is compatible with the interpretation that they represent a fast, final phase of disc evolution that proceeds from the inside out \citep{KenyonHartmann95}. It has to be remarked that, despite the images confirm large cavities both in the sub-mm and in the micron-sized dust, gas is in many cases still present. Indeed, some of these transitional discs present mass accretion rates of order $10^{-8} \ \mathrm{M_\odot yr^{-1}}$, comparable with that of classical T-Tauri stars, indicating that a substantial reservoir of gas is still present near the central object. Even more remarkably, it has to be noted that such mass accretion rates are not a general feature of all transition discs, and that there is instead a huge range of variation in the observed sample.

To explain the presence of this class of discs, many physical processes have been invoked, including grain growth \citep{2005A&A...434..971D}, photoevaporation \citep{UVswitch,AlexanderModels} and planet formation \citep{1999Natur.402..633A, 2003MNRAS.342...79R}. However, up to now none of these processes alone have been shown to be sufficient to explain all observations.

Growing through collisions, dust particles can indeed become large enough to become essentially invisible to observations. However, while models of grain growth are able to reproduce the observed dips in their infrared SEDs of transitional discs, they predict that the dust should still be visible at millimetric wavelengths, in contrast with what is found in observations \citep{TilNoGrainGrowth}.

The presence of a giant planet embedded in the disc is able to open a gap, and acts like a dam, stopping the inflow of matter from the outer disc reservoir. However, the dam is porous, and while the surface density in the inner disc is lowered and the mass accretion rate reduced, the material can still flow towards its way to the star \citep[e.g.][]{AccrThroughGap}. To reconcile this with observations, it is necessary to find the right combination of parameters that makes the inner disc optically thin, while still allowing a sensible mass accretion rate. As showed by \citet{Zhu2011}, theoretical calculations predict that one single planet is not able to perturb enough the surface density of the inner disc, and multiple accreting planets are required to open a gap of a size compatible with what is found in observations. This reduces however the mass accretion rates onto the star as well as the surface density. \citet{Zhu2011} conclude that even in the case of multiple planets it is not possible to interpret discs that exhibit a large hole size together with a high mass accretion rate onto the star.

Photoevaporation is the process through which high energy radiation (from the central star or from the environment) thermally drives a wind from the disc. The mass-loss rates depend on the detailed physics of the radiation field. Including the effect of EUV radiation from the central star, \citet{UVswitch} and \citet{AlexanderEvol} showed that the coupled evolution of a viscously evolving disc with the presence of a photoevaporative wind is able to open a gap in the inner disc when the mass accretion rate through the disc becomes comparable with the mass-loss rate of the wind. Owen et al (2012) further argue that the dust in the inner disc rapidly drifts onto the star \citep{2007MNRAS.375..500A} -- on a timescale of $\sim 10^3$ years--, the gas drains on its (much longer) viscous time-scale ($\sim 10^5$ years). The result is a disc that exhibits an inner dust cavity (hence a dip in the mid-infrared emission), and an inner gas disc that is still draining, producing a still measurable mass accretion signature. Much progress has been made in computing detailed mass-loss rates from the photoevaporative wind. This is crucial to determine the mass-accretion rate at which the wind is able to open a gap, thus determining the age of the disc and the properties of the resulting transition disc. In particular recent models have included in the calculation FUV and X-ray radiation \citep{2009ApJ...690.1539G,Ercolano2009,Owen11Models}.

In particular, \citet{Owen11Models} compared the statistics of transitional discs with evolution models including X-ray photoevaporation from the central star, showing that it is indeed possible to explain a large number of observed objects with photoevaporation alone. This model however still failed to reproduce the class of transitional discs with large holes and large mass accretion rates, due to the fact that, by the time photoevaporation has carved a large enough hole in the outer disc, the mass reservoir of the inner disc has dropped so much that no mass accretion rate is detectable anymore.

A possible scenario, as  suggested by \citet{2012MNRAS.426L..96O}, is that ``transitional'' discs are not a homogenous class, indicating that different physical processes may be at work, and there may be different paths to transitional disc formation, depending on which of these physical mechanisms is dominant. If this is the case, we expect that in some cases there may not be a single dominant process, and it may be the interplay among several of them that leads to a given transitional disc formation.

Along this route, the goal of this paper is to study if the combination of photoevaporation and planet formation, which have been up to now studied separately, can indeed help in interpreting the puzzling population of accreting transitional discs. By reducing the surface density and the mass accretion rate in the inner disc, we expect that the presence of a planet is able to trigger the opening of a gap by photoevaporation at early times. We call this process planet induced photoevaporation (PIPE). To investigate this scenario, we make use of hydrodynamics models of a photoevaporating disc with a giant planet embedded. Our purpose is to assess how the presence of a planet affects the clearing of the disc by X-ray photoevaporation.

%-Recently, Zhu combined the effect of a planet embedded in a disc with dust evolution (diffusion + drift, check exactly what; have to check also the one with multiple planets). Along the same line, we study here the combined effect of photoevaporation and planet formation. A giant planet is able to open a gap in a disc and to reduce the surface density and the mass accretion rate in the inner disc, therefore allowing photoevaporation to open a gap at early times, clearing the disc from inside out, while the inner disc is still draining onto the star (therefore showing higher mass accretion rates for a given hole size as compared to the planetless case). We call this  process PIPE (planet induced photoevaporation). The question we want to ask is whether these two combined effects can account for the population of discs with large holes and accretion rates.

This paper is structured as follows. In section 2 we present the numerical method we used and the results we obtained. In section 3 we discuss the results and in section 4 we draw our conclusions.

\section{Numerical investigation}

\subsection{Methods}
We study the disc-planet interaction process by means of the 2D grid-based hydrodynamics code \textsc{fargo} \citep{Fargo}. The conditions at the time $t_0$ of the formation of the planet are provided by a 1D viscous evolution code, that takes care of evolving the disc from time $t=0$ to $t_0$. This allows to save computational resources when detailed evolution of the disc is not needed.

\subsubsection{Initial conditions (1D evolution)}
As initial conditions, we use the models of \citet{Owen11Models}. We include the effects of viscous evolution and X-ray photoevaporation. The evolution of the surface density of the disc is described by the following equation:
\begin{equation}
\frac{\partial \Sigma}{\partial t} = \frac{3}{R} \frac{\partial}{\partial R}\left[R^{1/2}\frac{\partial}{\partial R}\left(\nu \Sigma R^{1/2}\right)\right] - \dot{\Sigma}_\mathrm{w} (R),
\end{equation}
where $\Sigma$ is the surface density, $\dot{\Sigma}_\mathrm{w} (R)$ is the photoevaporation profile (as in the Appendix of \citealt{Owen12Theory}) and $\nu$ is the kinematical viscosity coefficient, which sets the magnitude of viscosity. We choose the same values of the parameters as in \citet{Owen11Models}, that we summarise here. We evaluate $\nu$ using the $\alpha$ prescription \citep{AlphaViscosity}: $\nu=\alpha c_s H$, where $c_s$ is the sound speed of the gas, $H$ is the vertical scale-height and $\alpha$ the dimensionless Shakura-Sunyaev parameter. In our models we set $\alpha=1.5 \times 10^{-3}$. The sound speed is a fixed function of radius, and is chosen to give a mildly-flaring disc (i.e., $H/R \propto R^{\mathbf{1}/4}$); the normalization is chosen so that at $1 \ \mathrm{AU}$ the aspect ratio $H/R=0.0333$. Our computational grid covers the range $[0.0025 \ \mathrm{AU}, 2500 \ \mathrm{AU}]$, and it is comprised of 1000 grid points. The mesh is uniform in a scaled variable $X \propto R^{1/2}$. Our viscous code uses a flux-conserving donor-cell scheme, implicit in time. Details about the implementation can be found in \citet{2010A&A...513A..79B}.

The initial surface density profile is given by:
\begin{equation}
\Sigma(R,0)=\frac{M_\mathrm{d}(0)}{2 \pi R R_1} \exp (-R/R_1),
\label{eq:sigma}
\end{equation}
where $M_\mathrm{d}(0)$ is the initial mass of the disc and $R_1$ a scale radius describing the exponential taper of the disc's outer region. We set a value of $R_1=18 \ \mathrm{AU}$ and an initial disc mass of $0.07 \mathrm{M_\odot}$.

For what concerns the photoevaporation profile, there are two parameters, the mass of the central star $M_\ast$ and the X-ray luminosity $L_X$. We chose $M_\ast=0.7 \ \mathrm{M_\odot}$, while we perform calculations with different values of the X-ray luminosity. We do runs with the median X-ray luminosity $L_X = 1.1 \times 10^{30} \ \mathrm{erg \ s^{-1}}$, that we evolve for $2 \ \mathrm{Myrs}$ before inserting the planet, and runs with a higher X-ray luminosity of $\log L_X = 30.8$, that we evolve for $0.65 \ \mathrm{Myr}$. These values for the planet formation time-scale do not come from a physical model, but rather were chosen to have a similar reasonable surface density profile at the moment of the planet formation. Since in the case of the high X-ray luminosity the evolution of the disc is faster due to the increased mass-loss rate, in that case we chose a smaller value for the age of disc at the moment of planet formation. In both cases the normalization at $1 \ \mathrm{AU}$ is approximately $500 \ \mathrm{g \ cm^{-2}}$, which is a factor 3-4 lower than the Minimum Mass Solar Nebula \citep{MMSN}. However, it should be noted that the power-law slope is $-1$, rather than $-3/2$, so that in the outer region the surface density is higher. The total mass in the disc at the moment of planet formation is approximately $25 \ \mathrm{M_{jup}}$ for the median X-ray luminosity, and $20$ for the high X-ray luminosity (the difference due to the disc being more spread out in the first case), which is higher than the Minimum Mass Solar Nebula. We note that according to current planet formation theories it is difficult, although still plausible \citep{2010Icar..209..616M}, to form a gas giant in the short time-scale used in the second case; therefore the high X-ray luminosity case should be regarded as a limiting one.

\subsubsection{\textsc{fargo} simulations}
At time $t=t_0$, we assume that a gas giant planet forms, and we use the output of the 1D code as input for the 2D \textsc{fargo} code. We assume that the formation happens on a time-scale fast enough so that we can switch from a 1D disc withouth a planet to a 2D disc with a planet. The code, which has been widely used in studies of protoplanetary discs, solves the equations of hydrodynamics through finite differences on a grid in cylindrical coordinates. \textsc{fargo} uses the same algorithms as the \textsc{zeus} code \citep{Zeus} for hydrodynamics, but employs a modified azimuthal transport technique that result in a smaller computational request for disc geometries. %except in the azimuthal advection. Rather than doing it in a single step, and be then severely constrained in the timestep size by the Courant condition, the so-called \textsc{fargo} algorithm is able to significantly speed up the computation, by allowing each annulus to rotate at its local mean azimuthal velocity. In this way the Courant condition on the timestep takes into account only the deviations from the mean azimuthal speed, permitting to use much longer timesteps. This results in a huge saving of CPU time. The algorithm has been extensively tested \citep{Fargotests,2012arXiv1208.3170K} and it is known to produce results in very good agreement with the ones of the standard azimuthal transport technique, at a much lower computational expense.

The code solves the coupled system of Navier-Stokes and continuity equation. We modified the continuity equation from the publicly available version of the code to include the effects of photoevaporation on the disc. The continuity equation now reads:
\begin{equation}
\frac{\partial \Sigma}{\partial t} + \nablab \cdot (\Sigma \mathbf{v}) = - \dot{\Sigma}_\mathrm{w} (R),
\end{equation}
where $\dot{\Sigma}_\mathrm{w}$ is the same mass-loss profile due to photoevaporation employed in the 1D evolution. A similar implementation was also used to study planet scattering in transitional discs by \citet{2012MNRAS.419..366M}.The removal of mass is done at the beginning of the hydro time-step. To be able to follow the evolution in time of the disc, we implemented a minimum density through the disc. Whenever the density becomes smaller than a given threshold, it is reset to the minimum allowed value. We use the dimensionless value of $10^{-8}$ for this threshold, which is about five orders of magnitude smaller than the initial value of the density in the disc at the planet location. For safety, we modified also the timestep condition, adding another criterion that does not permit to photoevaporation to remove in a single timestep more than a given fraction $f$ of the mass in a cell. We used for $f$ the value of $0.1$. However, we noted in our simulations that this condition is never relevant, and that the other usual conditions on the timestep are more restrictive.

\begin{figure*}
\includegraphics[ trim=2cm 0cm 2cm 0cm, width=\textwidth]{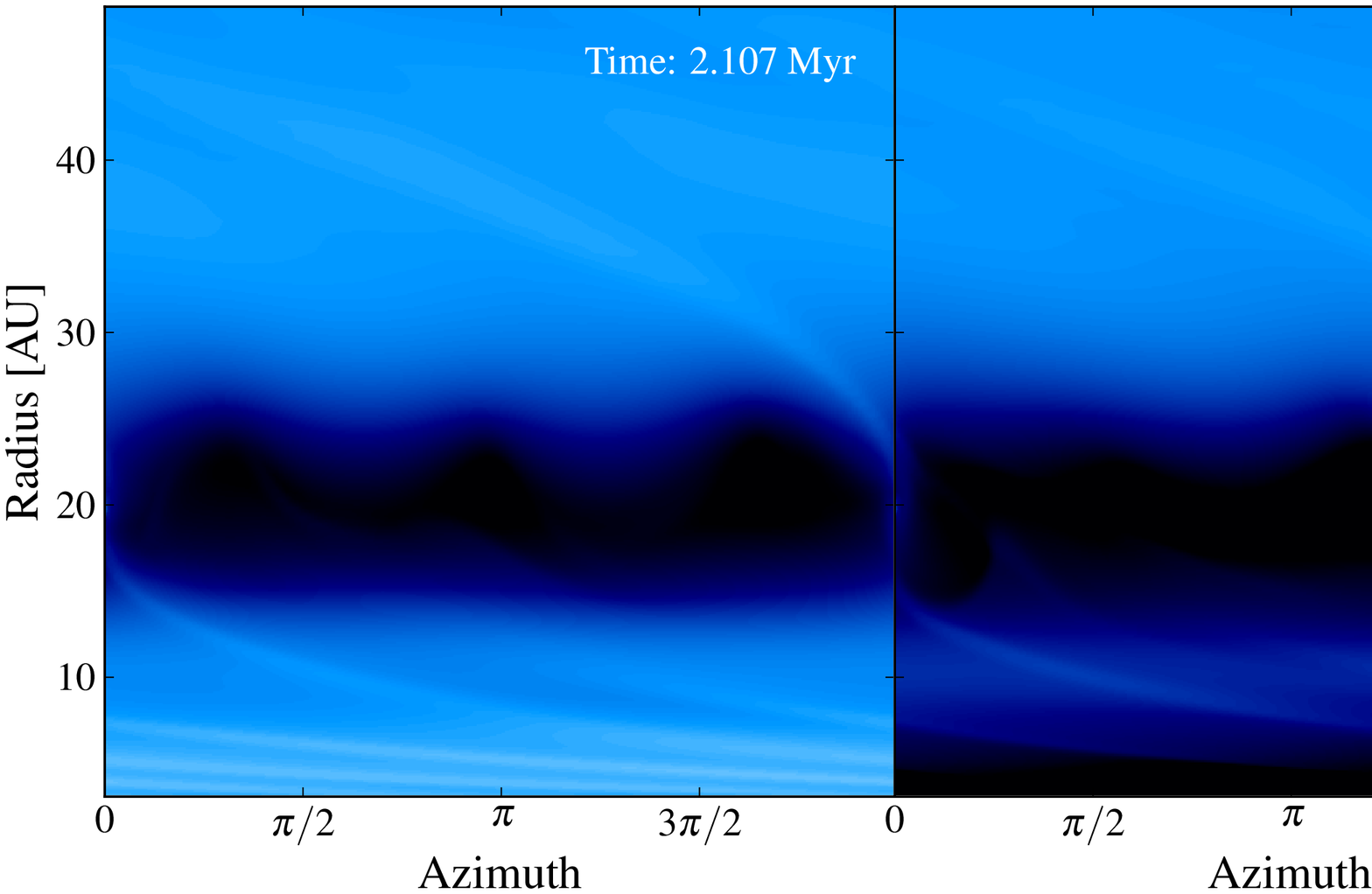}
\includegraphics[ trim=2cm 0cm 2cm 0cm, width=\textwidth]{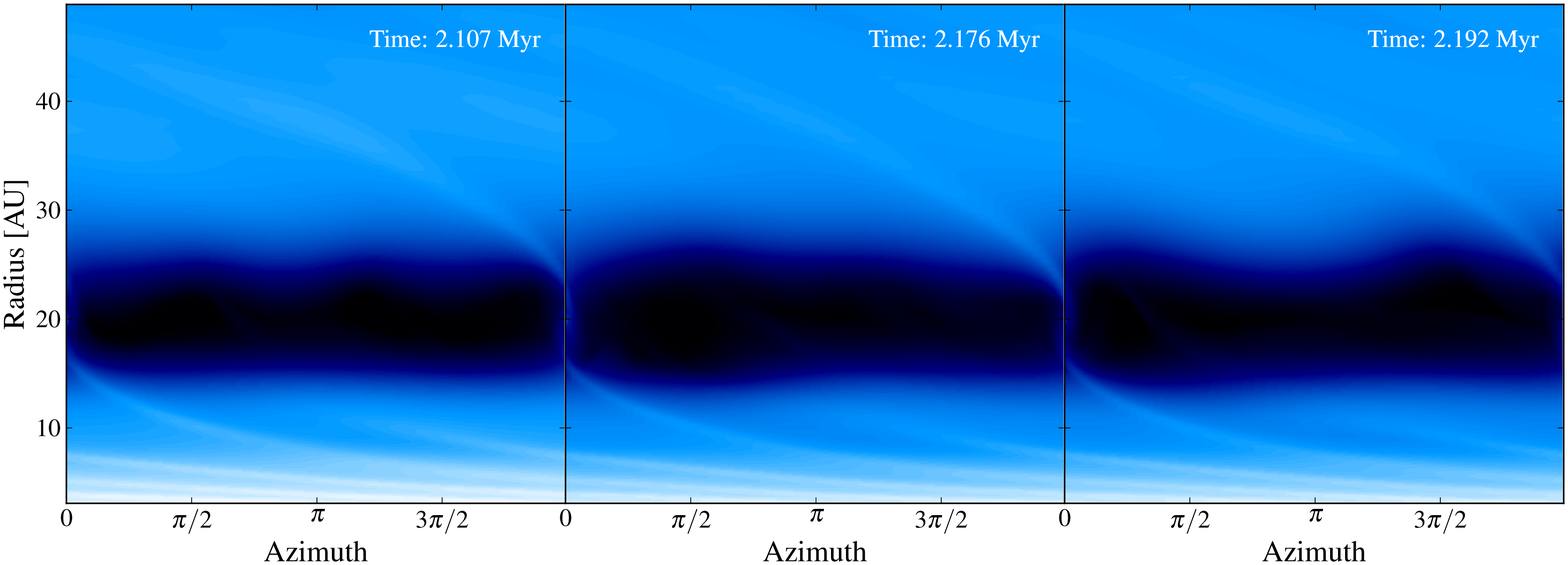}
\includegraphics [width=\textwidth]{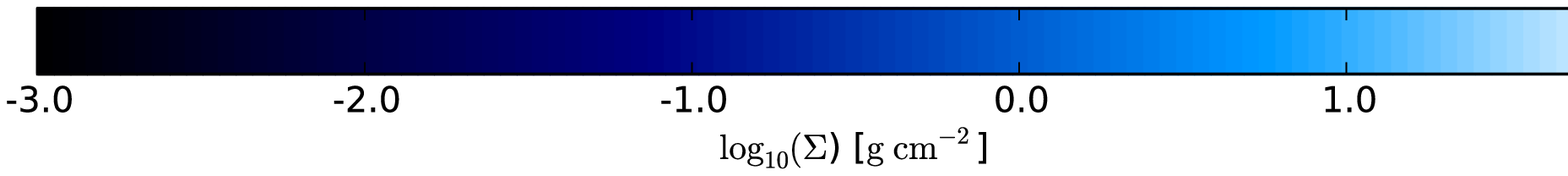}
\caption{Top row: surface density in the disc at three different times from simulation M20. Bottom row: same quantities for simulation M20o, which do not include photoevaporation. While in the first snapshot the inner disc is still there also in the case including photoevaporation, it is caught in the act of clearing in the second snapshot. Finally, we are left with a disc with the outer part only in the last snapshot. In the control simulation instead the inner disc is left.}
\label{fig:render}
\end{figure*}

We set the parameters to the same values as in the 1D evolution. We are however not able to resolve the whole disc in a 2d simulation, therefore we restrict ourselves in the range $[0.1 r_p, 10 r_p]$, where $r_p$ is the planet orbital radius. We employ a resolution of $n_\phi = 256$ cells in the azimuthal direction, uniformly distributed. The radial resolution is then chosen so that each cell is approximately square, which gives $n_r=188$ cells. While considerably higher resolution simulations can be found in the literature, we remark that here we are interested in the global evolution of the disc, rather than capturing some local property. Even if some feature is not properly resolved (such as the accretion streams onto the planet), this has little effect on the global evolution. For this reason similar studies of the disc evolution on a long timescale have employed a resolution similar to ours \citep[e.g.,][]{Zhu2011}. We have checked, running simulation M20 at double the resolution and comparing the result, that our results stay the same, with a maximum 10\% difference in the mass of the inner disc up to the moment of the disc clearing.

\begin{table*}
\begin{tabular}{ccccc}
\toprule
Run name & Orbital radius [AU] & $\log L_X$ & $t_0$ [Myr] & Notes\\
\midrule
M10 & 10 & 30.04 & 2 \\
M20 & 20 & 30.04 & 2 \\
M20m & 20 & 30.04 & 2 & migration\\
M20f & 20 & 30.04 & 2 & slower accretion\\
M20o & 20 & 30.04 & 2 & no photoevaporation\\
M30 & 30 & 30.04 & 2 \\
M40 & 40 & 30.04 & 2 \\
M50 & 50 & 30.04 & 2 \\
M20x & 20 & 30.8 & 0.65 \\
M40x & 40 & 30.8 & 0.65\\
\bottomrule
\end{tabular}
\caption{The table summarises the simulations run. The number in the name of the run specifies the position of the planet in AU; the ``x'' denotes the runs with high X-ray luminosity. In addition to the standard runs, which let the planet position and the X-ray luminosity vary, there also tree ``special runs'', with names M20m, M20f and M20o, respectively including migration, with a slower planetary accretion timescale employed and without photoevaporation. The initial planet mass is in every case $0.7 \ M_J$.}
\label{tab:sim}
\end{table*}

We consider a planet embedded in the disc with a mass $M_p=10^{-3} M_\ast = 0.7 \ M_J$, and we let the orbital radius of the planet vary from 10 to 50 AU. The planet is not allowed to migrate, although we also performed a run in which we include this effect. In the simulation in which we included also migration, we switched it on only after 500 orbits to restrict to the case of Type II migration and exclude Type I migration. To reduce artefacts in the solutions due to a sudden insertion of the planet, we slowly increase its mass during the first 100 orbits, using the taper function provided in the publicly available version of \textsc{fargo}. We use open inner boundary conditions and reflecting outer boundary (which is the default in FARGO. To test the impact of this choice, we implemented an outer boundary condition, finding no significant difference). We use planetary accretion as prescribed by \citet{Kley99} using the publicly available implementation in FARGO. At each timestep, a fraction of the material inside the Hill radius of the planet is removed and accreted onto the planet. The fraction is controlled by a free parameter, $f$, that represents the inverse of the accretion timescale in dynamical units. We use $f=1$ for our standard model, although we test also a situation with a 10 times slower accretion timescale. Table \ref{tab:sim} summarises the simulations run.

Lastly, it should be noted that the inclusion of a photoevaporation profile breaks down the degeneracy of the dimensionless units used by \textsc{fargo}. In the pure hydrodynamical case, we have one free mass scale and the results can then be scaled to different central star masses. This is no longer the case including photoevaporation, because $\dot{\Sigma}_\mathrm{w}$ depends on the (physical) radius in the disc and on the mass of the star.

\subsection{Results}

\subsubsection{Qualitative picture}
\label{sec:qualitative}

The surface density in the disc at three different times from runs M20 and M20o is plotted in figure \ref{fig:render}. In the left panel, at an age of approximately $2.1$ Myrs, the dynamical gap cleared by the planet is evident, but photoevaporation has not yet started to clear the disc. The surface density at this stage is very similar to a control run without photoevaporation, the most notable differences are visible at the gap edges. The planet acts like a dam for the viscous flow, reducing the mass accretion rate in the inner part of the disc. This permits photoevaporation to take over, and clear the inner disc, as can be seen in the intermediate step in which photoevaporation is clearing the disc from inside out. Finally, we are left only with the outer disc, when the disc is approximately $2.2$ Myrs old. It should be noted that the clearing of the inner disc is due to the combined effect of photoevaporation and planet formation. In the control simulation M20o without photoevaporation, the mass of the inner disc is reduced, but not cleared completely, as can be seen from the bottom row of figure \ref{fig:render}. On the other hand, with photoevaporation alone the disc would have cleared on a longer time-scale, after $3.3$ Myrs years. It can be concluded that PIPE, i.e. the combined effects of photoevaporation and planet formation, is able to clear the inner disc at earlier times, that is, stars with giant planets will dissipate their inner discs quicker.

\begin{figure}
\includegraphics[width=\columnwidth]{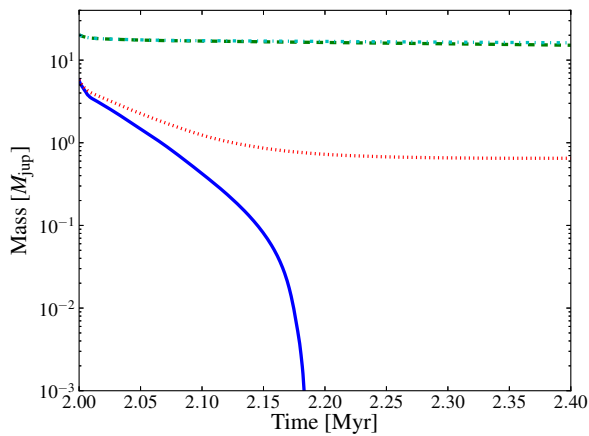}
\caption{Solid blue line: mass of the inner disc (i.e., inside the planet orbital radius) with photoevaporation (run M20). Dotted red line: mass of the inner disc without photoevaporation (run M20o). Dashed green line: mass of the outer disc (i.e., outside the planet orbital radius) with photoevaporation (run M20). Dotted-dashed cyan line: mass of the outer disc without photoevaporation (run M20o). In the run M20 with photoevaporation, the inner disc is rapidly dissipated, while in the control run M20o it reaches a sort of steady-state value.}
\label{fig:massdisc}
\end{figure}

Figure \ref{fig:massdisc} shows the masses of the outer and inner disc (i.e., at radii larger and smaller than the planet orbital radius) as a function of time. For reference we included also the result of the control run M20o without photoevaporation. While they start from the same initial value, the difference accumulates in time, and when photoevaporation becomes important it rapidly dissipates the inner disc. On the contrary, in the control run without photoevaporation the mass of the inner disc reaches some kind of steady state value, slightly smaller than a Jupiter mass. Due to our use of a floor density, after the clearing of the inner disc there is still a non-zero mass inside the orbit of the planet, with a value that is around $10^{-4} \ M_\mathrm{jup}$ (out of scale in the figure). It should be noted also that there is little difference in the mass of the outer disc between the two runs, and that the final value is around $15 \ \mathrm{M_{jup}}$.

\begin{figure}
\includegraphics[width=\columnwidth]{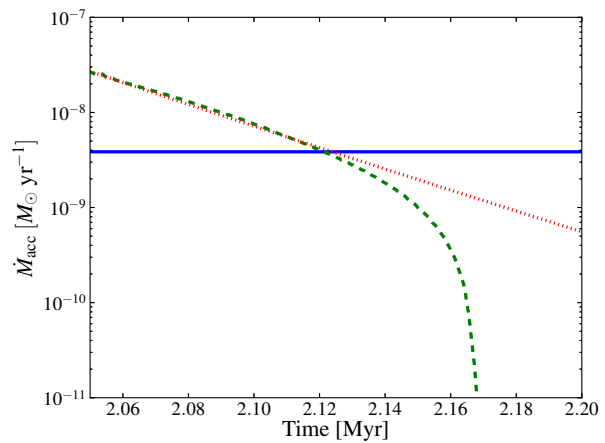}
\caption{Green dashed line: mass accretion rate at the inner boundary of the grid as a function of time for simulation M20. For reference, the blue solid horizontal line shows the mass-loss rate due to photoevaporation in the inner disc, while the red dotted line is a straight line added as a visual aid to distinguish the moment of the inner disc clearing. The change in the slope of the mass accretion rate corresponds to the inner disc clearing. This clearing happens when the mass accretion rate has dropped below some factor of the mass-loss rate. }
\label{fig:mdotdisc}
\end{figure}

\begin{figure*}
\includegraphics[width=\columnwidth]{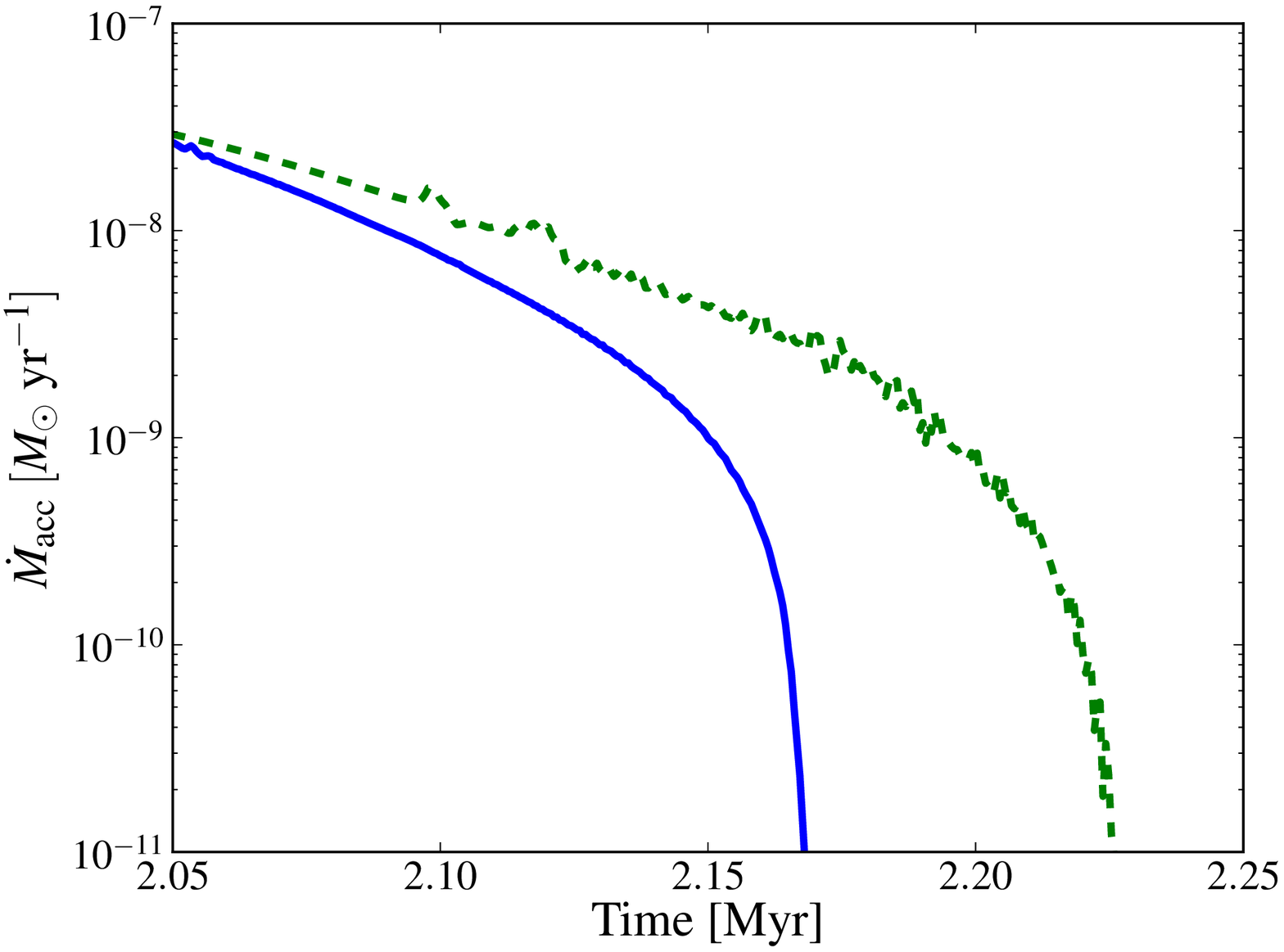}
\includegraphics[width=\columnwidth]{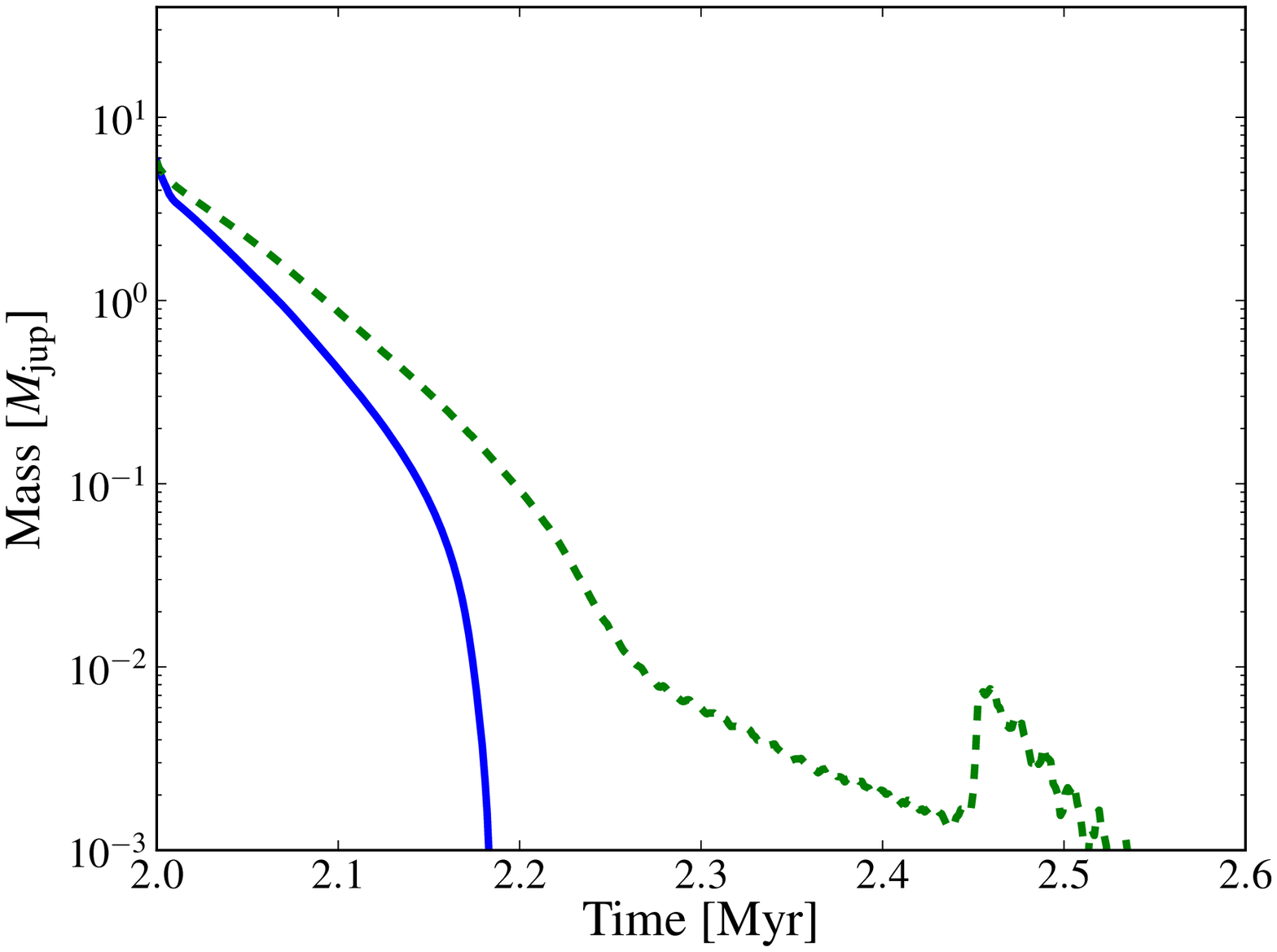}
\caption{Left: comparison of the mass accretion rate at the inner boundary as a function of time, varying the planetary accretion timescale. The blue solid line is for the standard value ($f=1$), run M20, and the green dashed line for the slow accretion ($f=10$), run M20f. Right: comparison of the mass of the inner disc as a function of time. The lines have the same style as in the left panel.}
\label{fig:accrcomparison}
\end{figure*}

In figure \ref{fig:mdotdisc} we plot the mass accretion rate at the inner boundary of the grid. The change in the slope corresponds to the moment at which the inner disc starts to clear. In the figure we overplot also, as a horizontal line, the mass-loss rate due to photoevaporation (considering the inner disc only). It can be seen that only once the mass accretion rate has dropped below some factor of the mass-loss rate the clearing begins. The qualitative understanding of disc clearing is thus not really different from the evolution of a disc without a planet, namely the two-timescale behaviour (as in the UV switch model): for most of its lifetime the evolution of the disc is driven by viscosity, and only when the mass accretion rate drops below the mass-loss rate due to photoevaporation the clearing begins. The role of the planet is that of accelerating this process, acting like a dam that reduces the mass accretion rate in the inner disc. We remark that, since the inner disc spends most of its lifetime with a mass accretion rate that is higher than the mass-loss rate, the total mass of the disc that has been lost through accretion is greater than the one carried away by photoevaporation. Thus, it is the accretion that does most of the ``dirty job'' of dissipating the disc, and photoevaporation only contributes in the last step of the removal. For what concerns planetary accretion, we have checked that most of the mass that ends up on the planet comes from the outer disc. This does not mean that planetary accretion does not play a role in the dispersal of the inner disc. Indeed, it controls the porosity of the planetary dam, fixing the amount of the viscous flow from the outer disc that is intercepted. This effect makes the mass of the planet increase by a factor 6 at the moment of disc dispersal. However, the direct effect, namely the mass accreted by the planet directly from the inner disc, is little compared with the mass leaving the grid from the inner boundary, apart for a small initial transient in which the planet is accreting from the region that will be dynamically cleared.

%Figure blah shows the sources of variation of the mass of the inner disc. We stress that the mass of the inner disc is not entirely carried away by photoevaporation. Since the initial mass is blah and the rate is blah, this would take blah time, that is much longer than what observed in the simulation. Rather, it is accretion onto the star that dissipates most of the mass of the disc. Only when the mass accretion rate in the disc has dropped below some treshold rate, photoevaporation will take over and clear the disc. 

%This can be seen in figure blah, where we plot the mass accretion rate as a function of radius for run xxx at time xxx (note: can also plot like sigma*vr, could be less noisy). The dam effect of the planet is evident, with the mass accretion rate in the inner disc that has dropped of a factor f. 

\subsubsection{Effect of planet accretion timescale}

This qualitative picture implies that a crucial parameter for estimating the impact of the presence of a planet in the disc is the porosity of the dam. We notice that in a 1D simulation the planet acts as a complete dam, and no filtering is possible. Thus, one has to insert by parametrization the porosity of the dam \citep[e.g.,][]{2009ApJ...704..989A}, in a way that does not conserve angular momentum. This effect is better accounted for in 2D simulations, where the transport of angular momentum across the gap is treated in a realistic fashion, although there is still a dependence on the planet accretion time-scale. To test how robust our results are with respect to the variation of this parameter, we run a simulation with a 10 times higher value (i.e., the planet accretes 10 times slower), $f=10$, that we call M20f. In the left panel of figure \ref{fig:accrcomparison} we propose a comparison between the mass accretion rate at the inner boundary of the grid of this simulation with the standard one. The inner disc is indeed cleared at later times, since we need to wait more for the mass accretion rate to drop, due to the higher porosity of the dam. However, it should be noted that the difference is not so dramatic as one could expect by such a big variation in the timescale. This is consistent with what other authors have found, namely that the process of planetary accretion is not dramatically dependant on the value of this parameter \citep{Kley99}. What cannot be seen from this plot however is that the way in which the disc clears is different. This can be seen from the mass of the inner disc versus time comparison in the right panel of figure \ref{fig:accrcomparison}. We can see that with a reduced planetary accretion time-scale not only the clearing happens at later times, but it is also slower. Visual inspection of the surface density distribution shows that an inner ring of material is left just inside the orbit of the planet, which is only slowly eroded. We interpret this as due to the fact that, because of the higher mass accretion rate through the planetary gap, photoevaporation has a much more difficult job at removing this part of the disc, being continuously replenished from the outer disc. This shows the complex structures that may be formed due to the interplay between the different physical processes acting in the disc.

\subsubsection{Varying planet position}

\begin{figure}
\includegraphics[width=\columnwidth]{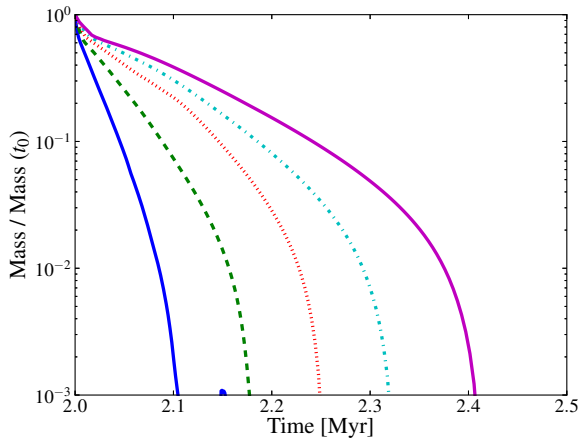}
\caption{Comparison of the mass of the inner disc normalized to the initial value for simulations with different position of the planet, varying from 10 to 50 AU (runs M10-M50). The further the planet, the slower the clearing of the disc.}
\label{fig:massinnerdiscposition}
\end{figure}

\begin{figure}
\includegraphics[width=\columnwidth]{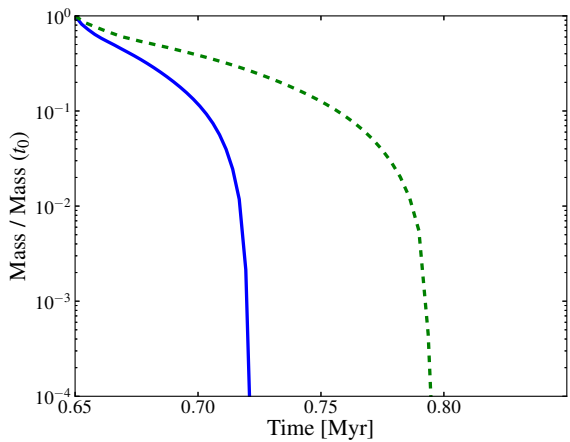}
\caption{Mass of the inner disc, normalized to the initial value, as a function of time, for simulations with the high value of the X-ray luminosity with positions of the planet of 20 and 40 AU, runs M20x (solid blue line) and M40x (dashed green line).}
\label{fig:massinerdischighX}
\end{figure}

In the simulation shown so far the planet has a semi-major axis of $20 \ \mathrm{AU}$. To find how the inner disc clearing depends on the planet position, we ran other simulations with different values of this physical parameter. Figure \ref{fig:massinnerdiscposition} shows a comparison in the mass of the inner disc (normalized to the initial value) as a function of time for the different simulations run, with the planet position varying from 10 to 50 AU (runs M10-M50). The further out the planet is, the slower the process of dispersal. This is expected, since more time is required for a further out planet to reduce the mass accretion rate near to the star. However, we can see that in all cases the disc dispersal is considerably faster than without the presence of a planet in the disc. A fit to the initial surface density profile with equation \ref{eq:sigma} gives for $R_1$ a value of approximately 80 AU, so that for the used values of the planet position the planet is able to cut most of the disc mass reservoir from the inner radii. From this argument we expect that a planet further out than $R_1$, if able to form, would not have a significant impact on the disc lifetime.

\subsubsection{Varying X-ray luminosity}

As a limiting case, we run simulations with a high value of X-ray luminosity, equal to logLx=30.8. The disc is $0.65 \ \mathrm{Myr}$ old at the time when the giant planet is inserted. While this may sound like an unrealistically young age for planet formation, we note that the relative disc age can be obtained by the appropriate scaling of initial parameters and here we chose a value that gives a similar surface density normalization (although the disc is less spread, being younger) to the previous case. This shows how even quite a massive disc can be rapidly dispersed by the combined effect of X-ray photoevaporation and planet formation. %Given the big number of free parameters, that are poorly constrained by observations, it is the value of the surface density at the time of the formation of the planet that should be looked at, rather than the absolute age.

The evolution of the mass of the inner disc as a function of time, normalized to the initial value, is shown in figure \ref{fig:massinerdischighX} for different values of the planet initial position. Because of the higher mass-loss rates of photoevaporation, everything is happening quicker, with a dispersal that can happen so fast as less than $10^5$ years after the planet formation. Also in this case the further the planet, the slower the process of disc dispersal. Visual inspection of the images confirms the same qualitative behaviour we outlined in the previous section.

\subsubsection{Effect of migration}
\label{sec:migration}
\begin{figure*}
\includegraphics[width=\columnwidth]{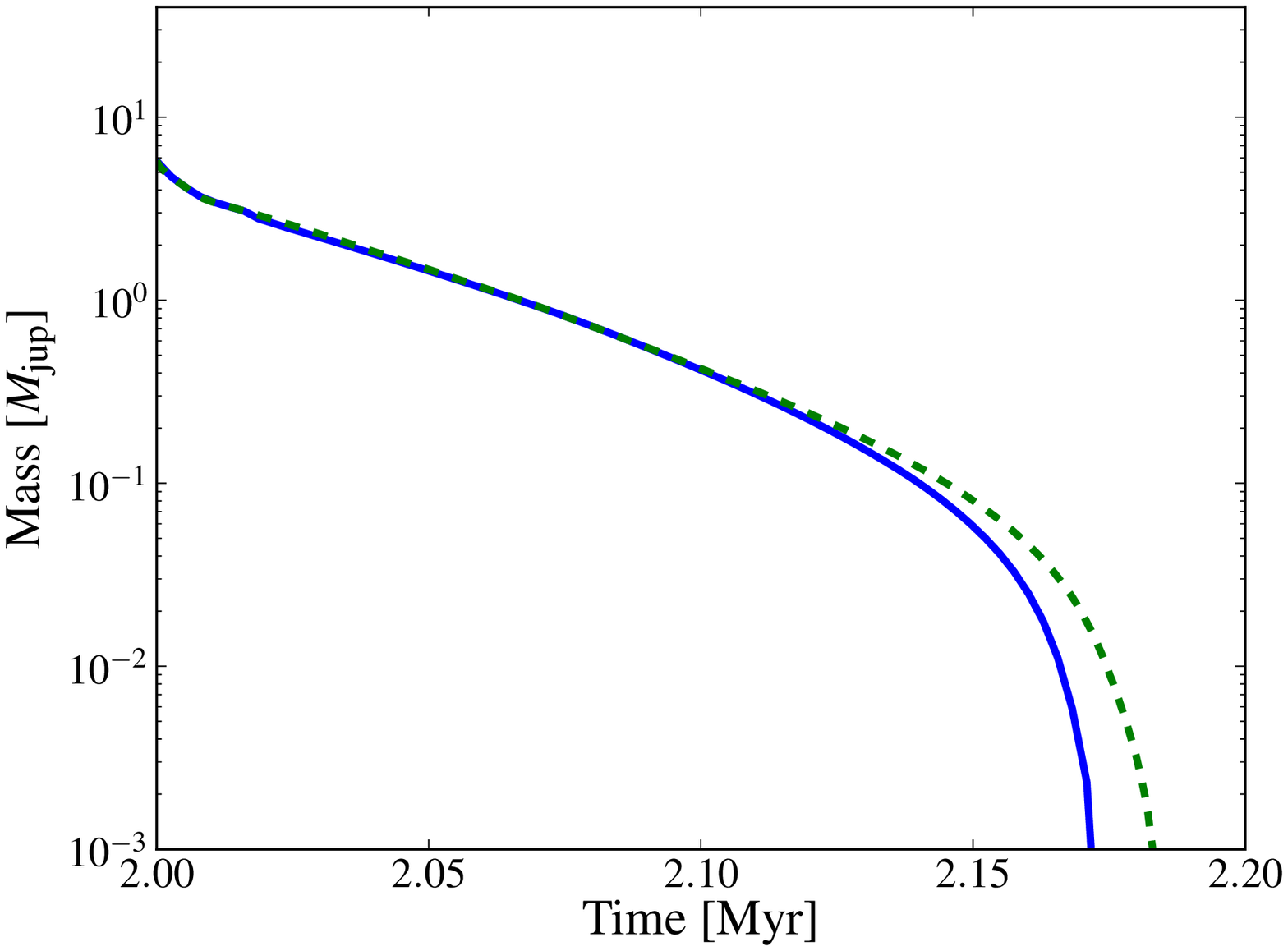}
\includegraphics[width=\columnwidth]{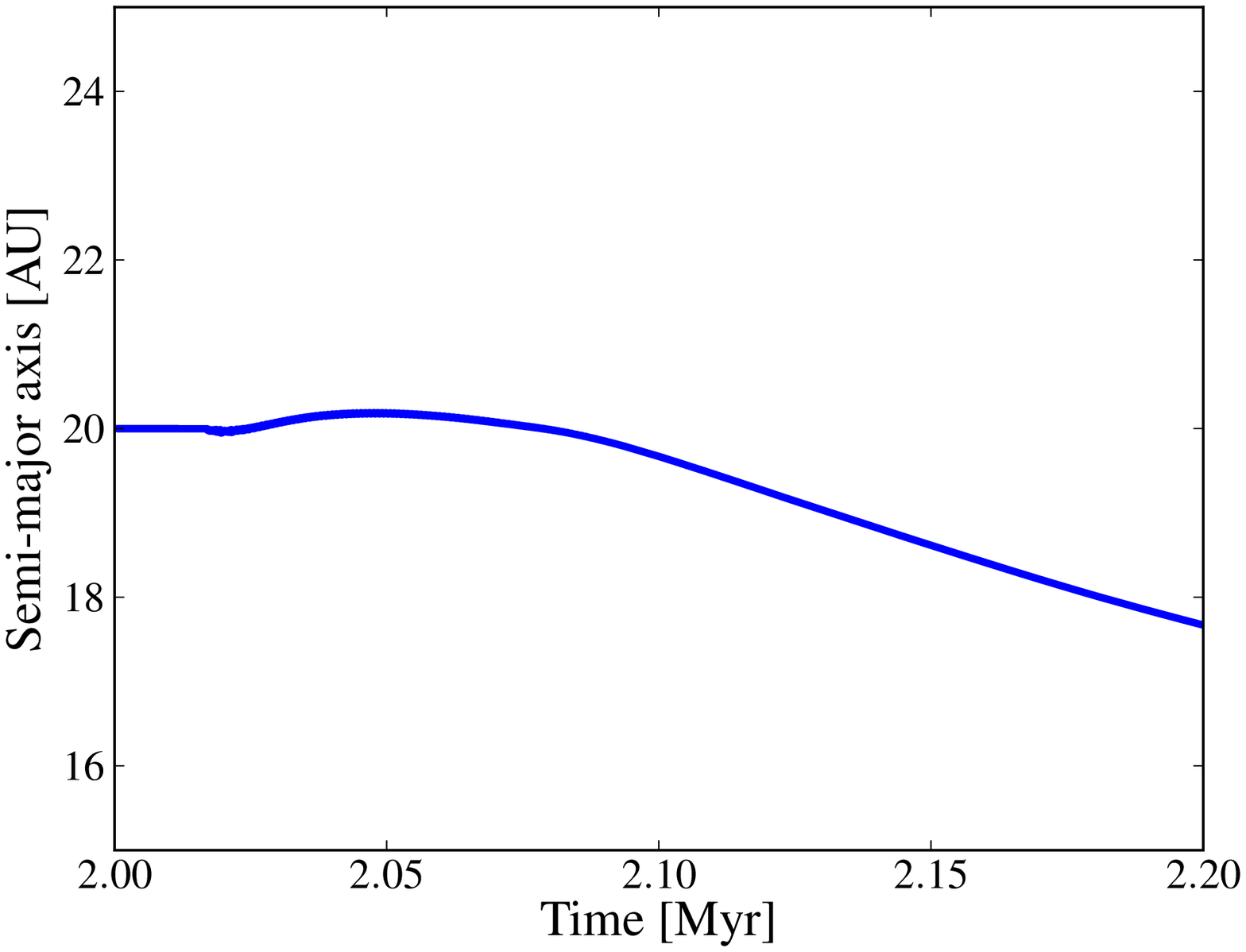}
\caption{Results from the simulation M20m including migration. Left panel: the mass of the inner disc as a function of time for simulation M20m (blue solid line). For reference has been plotted also the case without migration, run M20 (green dashed line). Right panel: semi-major axis of the planet as a function of time.}
\label{fig:migration}
\end{figure*}

To test the sensitivity of our results to migration, that we neglected so far, we run a simulation with migration included, M20m. The left panel of figure \ref{fig:migration} shows a comparison in the masses of the inner disc, showing that there is very little difference in the process of disc clearing. This is a consequence of the fact that the clearing proceeds from inside out, so that the beginning of the clearing is set by the properties of the disc in the inner portion, rather than in the neighbourhood of the planet that is affected by migration. The right panel of figure \ref{fig:migration} shows the semi-major axis a function of time, showing that the planet has not migrated much before the disc is dispersed, approximately 2 AU. It may be argued that this is a consequence of the initial conditions, namely that we took quite an evolved disc. Inserting the planet at earlier times would have allowed more time to migrate, and therefore to be more incisive in modifying the clearing. This is what has been studied by \citet{2009ApJ...704..989A} and \citet{2012MNRAS.422L..82A}, who coupled the disc evolution to the planet migration and studied the resulting planet distribution. The effect of more massive planets inserted in massive (self-gravitating) discs is the focus of a forthcoming study by Clarke et al. (in prep.), while in this work we focus on the study of less massive disc-planet systems.

\section{Discussion}
In the previous section we showed how PIPE is able to change the picture of disc dispersal with respect to photoevaporation or planet formation alone. The main finding is that discs with a planet are likely to be dispersed earlier than discs without. An interesting test of the PIPE scenario, is to compare its predictions with the observations of transitional disc, that are interpreted as discs caught in the act of clearing. In particular we study the $\dot{M}-R_\mathrm{hole}$ parameter space. For each of our models, the aim is to compute evolutionary tracks that can be plotted in this parameter space. In contrast with \citet{Owen11Models}, due to the increased computational cost, we are not able here to run a whole population synthesis; rather, we will be limited to comparing with individual datapoints. In particular, we wish to answer the question if transition discs with large holes and mass accretion rates can be accounted for by PIPE. A similar attempt has been made by \citet{Morishima12} through the modelling of X-ray photoevaporating discs with dead zones.

Unfortunately, our hydrodynamical modelling does not yield directly these parameters. In particular, the mass accretion rate the observations measure is the one onto the star, that we cannot resolve for numerical reasons. Therefore, we make two limiting assumptions to the modelling.

\begin{figure}
\includegraphics[width=\columnwidth]{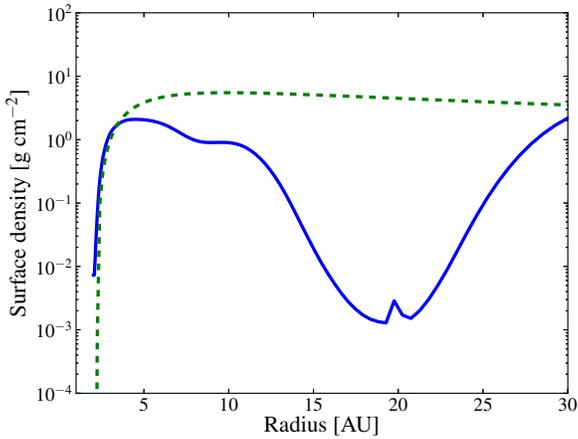}
\caption{Surface density as a function of radius just before the inner disc clearing starts in run M20 (blue solid line), compared with the surface density in the 1D calculations from \citet{Owen11Models} model with the median X-ray luminosity (green solid line). The surface density are quite similar near to the inner boundary, while at greater radii in the case with a planet the surface density profile is much flatter.}
\label{fig:densitysimilar}
\end{figure}

\begin{figure}
\includegraphics[width=\columnwidth]{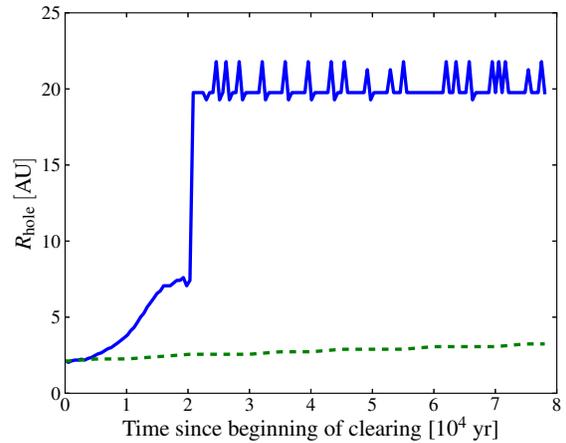}
\caption{Inner hole radius a function of time (since the clearing), for the PIPE M20 calculation (blue solid line) and the 1D model with the median X-ray luminosity (green dashed line) from \citet{Owen11Models}. The depletion of the surface density caused by the planet makes the opening of the hole much faster. The oscillations that can be seen in the radius of the hole in model M20 are a consequence of the oscillations at gap edges visible in figure \ref{fig:render}.}
\label{fig:rhole_t}
\end{figure}

\begin{figure}
\includegraphics[width=\columnwidth]{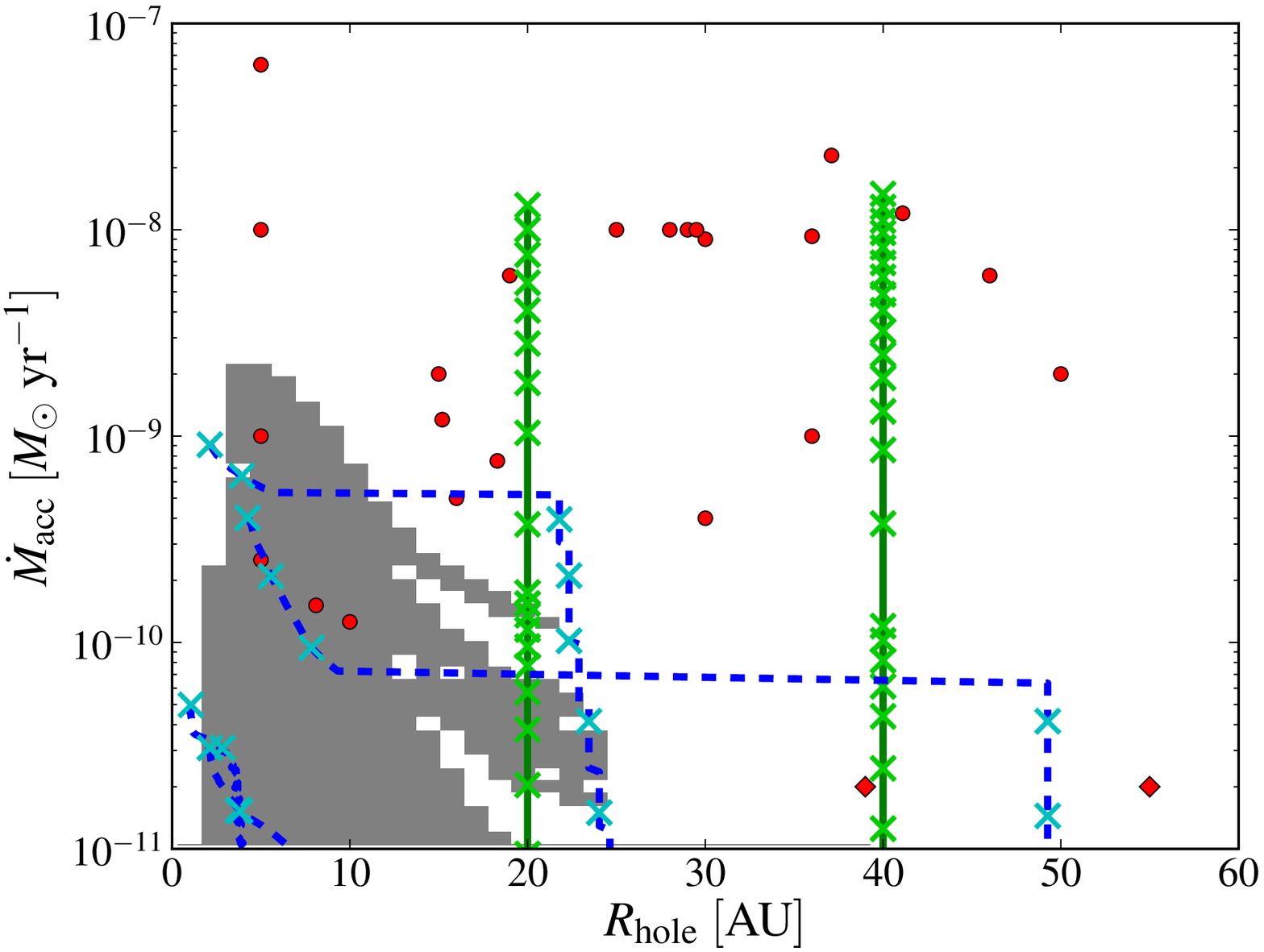}
\caption{$\dot{M}-R_\mathrm{hole}$ parameter space of transition discs. The grey area is the area permitted by X-ray photoevaporation alone, as in \citet{Owen11Models}. Points are the observed datapoints with high millimetre flux from \citet{2012MNRAS.426L..96O}. Diamonds are for disc without a detected mass accretion rate. Blue dashed tracks are for the first assumption (see text) and green continuous for the second; crosses on the tracks are plotted every $10^4$ years.}
\label{fig:parameterspace}
\end{figure}

In the first, conservative assumption, we assume that the X-ray photoevaporation driven clearing of the innermost disc proceeds as it would do in the absence of a planet, opening a gap at approximately $1.5 \ \mathrm{AU}$. This gap is inside the region that we can resolve in the simulation. The disc will then exhibit two gaps, the dynamical one cleared by the planet and the one opened by photoevaporation, that divide the disc in three distinct regions (named A,B and C in the following for clarity). As in the photoevaporation case, we assume that the dust in the innermost disc A will drain very quickly ($\sim 10^3 \ \mathrm{yr}$) onto the star, so that region A will be invisible in observations. The disc will then look like a transition disc and the radius of the hole is set by the inner radius of disc B, that we can measure from the 2D simulation. While A is draining, there is still a measurable mass accretion rate. We assume that the dependence on time of the mass accretion rate onto the star is the same as in the 1D simulation, where we take as the beginning of clearing the time at which in the 1D simulation the radius of the hole is the inner boundary of our 2D grid. We note that the density structure in the region we can resolve is quite similar before the onset of clearing of the inner disc, as we show in figure \ref{fig:densitysimilar}. In this figure we compare the surface density from our FARGO simulation M20 just before the inner disc begins to clear with the results from the 1D run without the planet, at the time in which the radius of the hole is 2 AU (namely the inner radius of our 2D grid). The surface densities near the inner boundary are very similar, while at larger radii the density distribution is lower, due to the presence of the planet that makes this region devoid of gas. This motivates us to assume that the dependence with time of the mass accretion rate onto the star is the same as in the case without the planet, coming from the draining of the disc A onto the star, but now with a different $R_\mathrm{hole}$ dependence that is derived from the 2D simulation. In figure \ref{fig:rhole_t} we compare the $R_\mathrm{hole}$ dependence for model M20 and the corresponding model without a planet, i.e. \citet{Owen11Models} model with the median X-ray luminosity. We define the radius of the hole simply as the minimum radius where the azimuthally averaged surface density is above the imposed floor density (with an allowance factor of 10, due to the presence of numerical oscillations above the floor density). Because of the mentioned depletion of the surface density, the clearing of the hole is much faster in the case with photoevaporation. When compared to models that include X-ray photoevaporation alone, we thus expect PIPE models to yield higher mass accretion rates for the same hole size, or conversely larger holes at the same mass accretion rate. The time $t=0$ in the plot refers to the beginning of the clearing, and it is the one when in the 1D simulation the radius of the hole is $2 \ \mathrm{AU}$.  We plot our results in the $\dot{M}-R_\mathrm{hole}$ parameter space in figure \ref{fig:parameterspace} as the blue dashed tracks. The cyan crosses are plotted every $10^4$ years. Runs M20x and M40x are the ones that produce the higher mass accretion rates, while only runs M10 and M20 are visible from the runs with the median X-ray luminosity. The points in the plot are the observed datapoints that exhibit a high millimetre flux, as it is described in \citet{2012MNRAS.426L..96O}. The datapoints show a weak correlation between the hole radius and the mass accretion rate, which has not been explained yet. The grey area is the area permitted by X-ray photoevaporation alone, as in \citet{Owen11Models}. Using this conservative assumption (blue lines), one would then conclude that the PIPE scenario is indeed able to expand the parameter space with respect to photoevaporation alone, but that the mass accretion rates obtained are still too low compared with the observations.

The other limiting assumption, that is a best case scenario for us, is the one in which the planet is able to filter the dust, so that the disc would look like in transition, even if there is still a huge quantity of gas inside the orbit of the planet. In synthesis, by modifying the surface density profile of the disc, the planet creates a pressure maximum. Dust particles, flowing inward, are unable to cross this maximum, so that the dust is filtered out from the inner portion of the disc, that becomes invisible to observations. This simple picture is complicated by the other dust processes that happen simultaneously (coagulation, fragmentation, diffusion, ...). This process has been studied in detail by \citet{Rice2006}, \citet{Pinilla12} and \citet{Zhu2012}. In this case, we take the orbital radius of the planet as the radius of the hole. Before the clearing starts, we take the mass accretion rate at the inner boundary of the grid, starting from 500 orbits since the beginning of the simulation (to exclude initial oscillations), under the assumption that the mass accretion rate in the inner portion of the grid that we do not simulate is set by the outer part that we can resolve. Once the inner disc starts to clear, we take as in the previous case the mass accretion rate as predicted by the case without the planet. The resulting tracks in this case (plotted in green continuous in figure \ref{fig:parameterspace}) look like vertical lines, since there is little evolution of the radius during the simulation, and the mass accretion rate simply decreases. Using this second assumption, one would then conclude that, when the dust properties are taken into account, PIPE is able to explain discs with large holes and mass accretion rates. While up to now the theoretical effort has been to predict whether the appearance of a planet bearing disc resembles that of transitional disc, little work has been done in understanding the evolution of such systems. Photoevaporation could potentially give a reason for the shutting down of accretion, that is requested from the lack of observed datapoints in the region of the plot with large holes and low mass accretion rates. To explore this possibility, we plotted light green crosses along the tracks every $10^4$ years. There is not however a clear separation between the uppermost and the lower part of the track, with only a little jump that is not enough to account for the desert on the observed desert in the observed distribution. If this desert is real and not due to observational biases (e.g. discs with high mass accretion rates are also more massive), then a new mechanism must destroy the outer disc. Such a mechanism could be given by the ``thermal sweeping'' effect, an instability found in photoevaporating discs with an inner hole by \citet{Owen12Theory}, able to destroy the disc in the order of the dynamical timescale.  The forthcoming work of Clarke et al (in prep.) presents an alternative scenario based on the carving of mm-bright accreting transition discs by massive planetary companions embedded and migrating in self-gravitating discs. %Furthermore, PIPE is able to give an explanation to why the accretion at a certain point stops. In particular, we have seen in section \ref{sec:qualitative} how the mass accretion rate suddenly drops when the inner disc starts to clear. Our model thus predicts a very rapid``jump" from a high mass accretion rate to a low one, which could possibly explain the lack of observed datapoints in the region of the plot with big holes and low mass accretion rates. Moreover, the small number of sources with big holes and no detected mass accretion rates also points out that, if this picture is correct, another mechanism must be at work to destroy the outer disc that is left, otherwise we would be left with a population of long-lived relic transitional discs with no detectable mass accretion rate. Such a mechanism could be given by the ``thermal sweeping'' effect, an instability found in photoevaporating discs with an inner hole by \citet{Owen12Theory}, able to destroy the disc in the order of the dynamical timescale.

Our results suggest that the interplay between X-ray photoevaporation and planet formation is worth of further study. Our work also implies that planets are an important ingredient in the disc dispersal process, and should be included in the modelling. The ability of the upcoming ALMA facility will be hopefully be able to tell more about the gas structure in transitional discs, which is at the moment very poorly constrained. 

\subsection{Model limitations}

In this paper it has been assumed that the photoevaporation profile is the same as computed by \citet{Owen11Models}, who modelled the photoevaporative flow in a disc without a planet. Large scale 3D simulations that include both photoevaporation and a planet embedded are being carried out and will be presented in a forthcoming paper (Rosotti et al, in prep.).

The absolute timescales discussed in the paper should be
taken with caution as they strongly depend on a number of
simplifying assumptions. First of all, a source of uncertainty 
comes from the unknown mechanism for the angular momentum 
transport. In this work we have considered a simple $\alpha$-disc 
and assumed a constant value of $\alpha$. The assumption of
a constant $\alpha$-value is common in many theoretical studies of discs
\citep[e.g.][]{AlexanderEvol,2009ApJ...705.1237G,Owen11Models}, and is
motivated by a lack of strong observational constraints and by simplicity, 
necessary to isolate the effect of the other processes that affect 
the evolution of the disc. However, no known theory of angular 
momentum transport predicts a constant value for $\alpha$ \citep{Armitage2011}. 
Therefore, such an assumption should be regarded more as a matter of 
convenience. Studies addressing the issue of how a physical
model for the viscosity, e.g. the layered accretion model proposed
by \citet{Gammie96}, changes the evolution of the disc have been conducted by, for example, \citet{2002MNRAS.334..248A,Morishima12}. The layered accretion model predicts the formation of dead zones in the disc, where the viscosity is significantly lower than in the active parts of the disc. This is due to the insufficient ionization level in the cold midplane of the disc, which is not enough to couple the gas with the magnetic field and trigger the magnetorotational instability (MRI). The accretion continues in a thin layer, where the ionizing radiation (X-rays from the star and cosmic rays) are able to penetrate. The net effect, once vertically averaged, is to create a zone with a reduced viscosity. Employing a different model of photoevaporation than the one used in this paper, \citet{2002MNRAS.334..248A} modelled the combined evolution of a giant planet migrating in a layered disc. They find that the main effect of layered accretion is to slow down planetary migration due to the longer viscous timescale. However, as showed in section \ref{sec:migration}, migration is already relatively unimportant for our results, so this is probably a second-order effect. \citet{Morishima12} found that the interplay between dead zones and photoevaporation can also be a possible route for the formation of transition discs with large holes and large mass accretion rates. Their results predict that, near the outer edge of a dead zone, the radial motion of gas is directed outwards, while it is inwards inside the dead zone region. This effect, once coupled with photoevaporation, may be able to open a gap in the disc at the outer edge of the dead zone, around $~40 \ \mathrm{AU}$ in their calculations. The dead zone in the inner disc will survive for a long time after the opening of the gap, so that $\dot{M}$ remains high even after the gap opens. Such considerations may also apply to the systems explored in this work. It is however difficult to predict the effect of a varying viscosity induced by the presence of a dead zone on the disc clearing timescales calculated here. The location and extent of the supposed dead-zone with respect to the planet will have a strong influence on how the dispersal is affected. Even the direction of the feedback from dead-zone formation  is difficult to predict. On the one hand, a dead zone reduces the mass accretion rates in the region, hence reducing the mass flows through a planet gap, yielding to a faster decoupling of the inner and outer disc. However if the dead-zone contains most of the disc mass, as in the \citet{Morishima12} calculations, its effect could also be one of stabilisation of the mass accretion rates in the inner disc, resulting in a slower draining of the latter. Clearly a focussed exploration of the relevant parameter space would be required in order to provide educated predictions of the interaction of dead-zones (or indeed variable viscosity) with PIPE, which is beyond the scope of this work. 
Further uncertainties in the absolute timescales are introduced by numerical limitations,
 in particular the underresolution of the inner boundary that influences the resulting inner
disc clearing timescales. Experiments have shown that the effect of
having an inner boundary at a larger radius is to produce a
higher mass accretion rate at the beginning of the simulation.
This makes the disc deplete faster, thus shortening the lifetime
of the disc. Therefore, while a faster disc evolution due to the
combined effect of photoevaporation and planet formation is
a robust prediction of PIPE (as can be seen by comparison of
figure \ref{fig:render}), absolute disc lifetimes may be longer than what presented
here. While this effect is irrelevant for the first assumption
presented here in the discussion, it could partially help in
explaining the mentioned desert in the observed transitional
disc population. In addition, as already said in the previous sections,
this numerical limitation precluded us the possibility of
following the evolution of the very inner disc, where photoevaporation
is opening a gap.

Finally, another possible improvement of the models presented here is adding the modelling of the dust, in order to be able to compare directly the outcome of the model with observations. This will be particularly important to compare with ALMA observations.

\section{Conclusions}
In this paper we presented results from 2D simulations of discs with giant planets embedded undergoing X-ray photoevaporation. Our results show that planet formation influences the process of disc dispersal by photoevaporation. The main consequences of planet formation induced photoevaporation (PIPE) can be summarized as follows:

\begin{enumerate}

\item by reducing the mass accretion flow onto the star, discs that form planets will be dispersed at earlier times than discs without by X-ray photoevaporation.

\item For what concerns transitional discs, PIPE is able to produce transition disc that for a given mass accretion rate have larger holes when compared to standard X-ray photoevaporation. However further modelling of the dust processes is needed to be able to fully exploit the observational consequences of this process.

\item Assuming that the planet is able to filter dust completely \citep{Rice2006,Pinilla12,Zhu2012}, large hole transition discs could be produced. PIPE may instigate the shutting down accretion; however, our simplified models cannot at present explain the observed desert in the population of transition disc with large holes and low mass accretion rates.

\end{enumerate}

\section*{Acknowledgements}
Giovanni Rosotti acknowledges the support of the International Max Planck Research School (IMPRS) and is grateful to CITA for its hospitality during part of this work. PJA acknowledges support from NASA under grant HST-AR-12814 awarded by the Space Telescope Science Institute, which is operated by the Association of Universities for Research in Astronomy, Inc., for NASA, under contact NAS 5-26555. We would like to thank Tilman Birnstiel for providing the viscous evolution code. We thank Giuseppe Lodato and Cathie Clarke for interesting discussions.

\bibliography{BiblioRosotti}{}
\bibliographystyle{mn2e}

\bsp

\label{lastpage}

\end{document}